\let\includefigures=\iffalse
%
\let\useblackboard=\iftrue
%
%
\newfam\black
\input harvmac
\useblackboard
\message{If you do not have msbm (blackboard bold) fonts,}
\message{change the option at the top of the tex file.}
\font\blackboard=msbm10
\font\blackboards=msbm7
\font\blackboardss=msbm5
\textfont\black=\blackboard
\scriptfont\black=\blackboards
\scriptscriptfont\black=\blackboardss
\def\Bbb#1{{\fam\black\relax#1}}
\else
\def\Bbb{\bf}
\fi

\def\Rb{{\Bbb R}}
\def\Zb{{\Bbb Z}}

\def\nl{\hfill\break}
\def\p{\partial}

\def\II{\relax{I\kern-.10em I}}
\def\IIa{{\II}a}

\def\p{{\partial}}

\def\build#1_#2^#3{\mathrel{
\mathop{\kern 0pt#1}\limits_{#2}^{#3}}}
\def\boxit#1#2{\setbox1=\hbox{\kern#1{#2}\kern#1}%
\dimen1=\ht1 \advance\dimen1 by #1 \dimen2=\dp1 \advance\dimen2 by #1
\setbox1=\hbox{\vrule height\dimen1 depth\dimen2\box1\vrule}%
\setbox1=\vbox{\hrule\box1\hrule}%
\advance\dimen1 by .4pt \ht1=\dimen1
\advance\dimen2 by .4pt \dp1=\dimen2 \box1\relax}
\Title{\vbox{\baselineskip12pt\hbox{hep-th/9711165}
\hbox{IHES/P/97/83}
\hbox{RU-97-95}
\hbox{QMW-PH-97-34}
\hbox{LPTENS 97/56}}}
{\vbox{
\centerline{D-branes and the Noncommutative Torus} }}
\centerline{Michael R. Douglas$^{1,2}$ and Chris Hull$^{3,4}$}
\medskip\centerline{$^1$ Institut des Hautes \'Etudes Scientifiques}
\centerline{Le Bois-Marie, Bures-sur-Yvette, 91440 France}
\medskip
\centerline{$^2$ Department of Physics and Astronomy}
\centerline{Rutgers University }
\centerline{Piscataway, NJ 08855--0849 USA}
\medskip
\centerline{$^3$ Physics Department, QMW }
\centerline {Mile End Road, London E1 4NS, U.K.}
\medskip
\centerline{$^4$ Laboratoire de Physique Th\' eorique, Ecole Normale Sup\' erieure, }
\centerline{
24 Rue Lhomond, 75231 Paris Cedex 05, France.}
\medskip
\centerline{\tt douglas@ihes.fr, c.m.hull@qmw.ac.uk.}
\bigskip
We show that in certain superstring compactifications,
gauge theories on noncommutative tori will naturally appear as
D-brane world-volume theories.
This gives strong evidence that they are well-defined quantum theories.
It also gives a physical derivation of the identification proposed
by Connes, Douglas and Schwarz of Matrix theory compactification
on the noncommutative torus with  M theory compactification
with constant background three-form tensor field.

\Date{November 1997}
%
\lref\bfss{T. Banks, W. Fischler, S. H. Shenker and L. Susskind,
Phys. Rev. D55 (1997) 5112-5128; hep-th/9610043.}
\lref\dhn{B. de Wit, J. Hoppe and H. Nicolai,
Nucl.Phys. {\bf B 305 [FS 23]} (1988) 545.}
\lref\town{P. Townsend, to appear in the Strings '97 proceedings.}
\lref\bst{E. Bergshoeff, E. Sezgin and P. K. Townsend,
Phys. Lett. 189B (1987) 75;
Ann. Phys. 185 (1988) 330.}
\lref\DLN{B. de Wit, M. L\"uscher and H. Nicolai,
Nucl.Phys. {\bf B 305 [FS 23]} (1988) 545.}
\lref\HT{C. Hull and P. K. Townsend}
\lref\Witone{E. Witten}
\lref\DLP{J.~Dai, R.~G.~Leigh and J.~Polchinski,
Mod. Phys. Lett. {\bf A4} (1989) 2073.}
\lref\Pol{J.~Polchinski, Phys.~Rev.~Lett.~{\bf 75} (1995) 4724-4727;
hep-th/9510017.}
\lref\susslc{L. Susskind, hep-th/9704080.}
\lref\dkps{M. R. Douglas, D. Kabat, P. Pouliot and S. Shenker,
Nucl. Phys. B485 (1997) 85-127; hep-th/9608024.}
\lref\doug{M. R. Douglas, to appear in the '97 Les Houches proceedings}
\lref\ikkt{N. Ishibashi, H. Kawai, Y. Kitazawa, A. Tsuchiya,
Nucl.Phys. B498 (1997) 467-491.}
\lref\grt{O. Ganor, S. Ramgoolam and W. Taylor,
Nucl. Phys. B492 (1997) 191-204; hep-th/9611202.}
\lref\suss{L. Susskind, hep-th/9611164.}
\lref\Banks{T. Banks, ``Matrix Theory,'' hep-th/9710231.}
\lref\cds{A. Connes, M. R. Douglas and A. Schwarz, 
``Noncommutative Geometry and Matrix Theory: Compactification on Tori,'' 
hep-th/9711162.}
\lref\connes{A. Connes, {\it Noncommutative Geometry,} Academic Press, 1994.}
\lref\connesym{A. Connes,
Comm. Math. Phys. 182 (1996) 155-176; hep-th/9603053.}
\lref\connesmath{A. Connes, C. R. Acad. Sci. Paris S\'er. A-B 290 (1980)
A599-A604; \nl
M. Pimsner and D. Voiculescu, J. Operator Theory 4 (1980) 93--118; \nl
A. Connes and M. Rieffel, ``Yang-Mills for noncommutative two-tori,''
in Operator Algebras and Mathematical Physics (Iowa City, Iowa, 1985),
pp. 237--266, Contemp. Math. Oper. Algebra. Math. Phys. 62,
AMS 1987; \nl
M. Rieffel, ``Projective modules over higher-dimensional
non-commutative tori,'' Can. J. Math, 40 (1988) 257--338.}
\lref\taylor{W. Taylor, Phys. Lett. B394 (1997) 283-287; hep-th/9611042.}
\lref\defquant{M. De Wilde and P. Lecomte, Lett. Math. Phys. 7 (1983) 487; \nl
B. V. Fedosov, J. Diff. Geom. 40 (1994) 213; \nl
M. Kontsevich, q-alg/9709040.}
\lref\dhn{B. de Wit, J. Hoppe and H. Nicolai,
Nucl. Phys. B305 [FS 23] (1988) 545.}
\lref\ncfields{C. Klim{\u c}ik, hep-th/9710153.}
\lref\hoppe{J. Hoppe, Phys. Lett. B250 (1990) 44.}
\lref\zfc{C. Zachos, D. Fairlie and T. Curtright, hep-th/9709042.}
\lref\baake{M. Baake, P. Reinicke and V. Rittenberg, J. Math Phys. 26 (1985)
1070; \nl
R. Flume, Ann. Phys. 164 (1985) 189; \nl
M. Claudson and M. B. Halpern; Nucl. Phys. B250 (1985) 689.}
\lref\tfour{
M. Rozali, Phys. Lett. B400 (1997) 260-264, hep-th/9702136; \nl
M. Berkooz, M. Rozali and N. Seiberg, hep-th/9704089.}
\lref\hack{F. Hacquebord and H. Verlinde, hep-th/9707179.}
\lref\townsend{P. K. Townsend, hep-th/9612121;
to appear in procs. of the ICTP June 1996 summer school.}
\lref\senseiberg{A. Sen, ``D$0$ Branes on $T^n$ and Matrix Theory,''
hep-th/9709220; \nl
N. Seiberg, ``Why is the Matrix Model Correct?'' hep-th/9710009.}
\lref\tdual{A. Giveon, M. Porrati and E. Rabinovici,
Phys.Rept. 244 (1994) 77-202, hep-th/9401139; and references there.}
%
It is often stated that in toroidal compactification in superstring theory,
there is a minimum radius for the torus, the string length.  Tori with
smaller radii can always be related to tori with larger radii by using
T-duality.

However, this picture is not correct in the presence of other background fields, as is clear from the following simple example.
Consider compactification on $T^2$
with a constant Neveu-Schwarz two-form field $B$.  
We quote the standard result \tdual: under simultaneous T-duality of
all coordinates, the combination $(G+B)_{ij}$ is inverted, 
where $G$ is the metric expressed in string units.
Consider a square
torus and take the limit 
\eqn\tlimit{
R_1=R_2\rightarrow 0; \qquad B \ne 0\ {\rm fixed}.
}
The T-dual torus has
$$G+B = {1\over R_1^2R_2^2 + B^2} 
\left(\matrix{R_2^2& -B\cr B& R_1^2}\right)
$$
and the T-dual radii go to zero as 
$\tilde R_1 = \tilde R_2 \sim R_1/B$.

Of course the full T-duality group is larger and includes an $SL(2,\Zb)$
acting on the upper half plane $B+iR_1R_2$ in the usual way.
For any point there exists
an $SL(2,\Zb)$ transformation which brings it to the fundamental
region with $R_1R_2\ge\sqrt{3}/2$, so that it would appear that there is 
again a minimal length.
However, the particular $SL(2,\Zb)$ transformation that maps back to the standard fundamental region depends on
$R_1R_2$ and varies wildly as we take the limit \tlimit.
In any situation involving objects which transform under $SL(2,\Zb)$, the
physics of the limit is not well described by making these transformations.

A better way to get at the limit is to T-dualize a single dimension.
This exchanges the complexified K\"ahler form $B+iR_1R_2$ and the complex
structure of the torus $\tau=R_2/R_1$, to produce a theory with fixed
volume and $B=0$, while the torus becomes highly anisotropic:
$\tau' = B+iR_1R_2$.  
Let us think of this torus as $\Rb^2$ quotiented by the two
translations $x\rightarrow x+1$ and $x\rightarrow x+\tau'$.
In the limit $R_1R_2\rightarrow 0$, the original
$T^2$ appears to degenerate to $\Rb\times S^1$ with coordinate 
$0\le x^1 < 1$ quotiented by an additional
translation $x^1\rightarrow x^1+B$.
The nature of such a quotient depends radically on whether $B$ is
rational or irrational; if we interpret it in the latter case naively
as a pointwise identification, it will not lead to a conventional
Hausdorff manifold.

This is exactly the situation which was the original motivation
for noncommutative geometry \connes, leading us to try to interpret the
resulting space as a noncommutative space.  Not only is this a sensible
thing to do, we will now give a physical argument which leads to
gauge theory on the noncommutative torus as defined in \connesmath. 

Consider a theory of $N$ D$0$-branes on the original torus.
After T-dualizing one dimension, 
this becomes a theory of $N$ D$1$-branes stretching
along the T-dualed dimension; let them be extended in $x^1$, and placed
at $x^2=0$.  The compactness of $x^2$ will be implemented by placing
images of the D$1$-branes at $x+nR_2$ for all integers $n$; in other
words extended in $x^1$ and at $x^2=nR_2$.\footnote*{
The periodicities after T duality are $\sqrt{V'/\Im\tau'}=1/R_1$
and $\sqrt{V'\Im \tau'}=R_2$.}

This is not a conventional $U(N)$ $1+1$-dimensional gauge theory,
because it contains additional light states.
A string winding about $x^2$ will take the shortest path between a
D1-brane and its image consistent with the boundary conditions on the D1-brane,
Dirichlet in $x^2$ and Neumann in $x^1$.
This means that the string will end at a right angle and follow a path
$(x^1,s)$ for $0\le s< R_2$.  Its length is $R_2$ and it is 
light, with mass $m \propto R_2$, so it must be kept
in the low energy theory. 

A string winding $w_2$ times will have mass
$m \propto w_2 R_2$, so we must consider fields with arbitrary dependence
both on $x^1$ and integral $w_2$ to get the complete low energy spectrum.
By the usual Fourier transform of $w_2$ to $x^2$, we could rewrite this as a
$U(N)$ $2+1$-dimensional gauge theory, in terms of fields 
$A_i(x^1,x^2)$.

However, this is not a conventional $U(N)$ $2+1$-dimensional gauge theory
either, because the open string which starts at $(x^1,0)$ will end
at $(x^1,w_2R_2)$ which is identified with
$(x^1-w_2 B/R_1,0)$.  The degrees of freedom do not live at points in 
$2+1$-dimensional space and thus the 
theory is not local in the conventional sense.

Still, it is not hard to guess what modification of the gauge theory is
induced by non-zero $B$.  When two open strings interact, the end of the
first must coincide with the beginning of the second.  We can express this
in the $(x^1,w_2)$ basis by an interaction of the form
$$\eqalign{
S_{int} &= \sum_{w_2,{w'}_2} \int dx^1 \phi_1(x^1,-w_2-w'_2) \phi_2(x^1,w_2) 
	\phi_3(x^1-Bw_2/R_1,w'_2) \cr
&= \sum_{w_2,{w'}_2} \int dx^1 \phi_1(x^1,-w_2-{w'}_2) \phi_2(x^1,w_2) 
	\exp\left(-{Bw_2\over R_1}{\p\over\p x^1}\right) \phi_3(x^1,{w'}_2) .
}$$
We then Fourier transform, and go to coordinates of unit radius
$\sigma_i\equiv R_i x^i$.
The interaction becomes
\eqn\interact{\eqalign{
S_{int} &= 
\int d\sigma^1 d\sigma^2\ \phi_1(\sigma^1,\sigma^2)  
	\exp\left({B\over 2\pi i}{\p\over\p {\sigma^1}'}
		{\p\over\p {\sigma^2}'}\right) 
\phi_2(\sigma^1,{\sigma^2}') \phi_3({\sigma^1}',\sigma^2)
		\bigg|_{{\sigma^i}'=\sigma^i}.
}}
But this is exactly the characteristic
interaction term in gauge theory on the noncommutative torus!
Without repeating all the details (given in \cds\ and other
works on this theory), the general connection on the noncommutative torus
is a sum $\nabla_i+A_i$ where $\nabla_i$ is a constant curvature connection  
and $A_i$ are are elements of the algebra defined by
$$Z^2 Z^1 = e^{2\pi iB}Z^1 Z^2.$$
One can be more concrete by choosing an identification 
$Z^i \sim e^{2\pi i\sigma^i}$;
then $\nabla_i = \p/\p \sigma^i$ and the
noncommutative multiplication is exactly realized by the differential
operator which appears in \interact.  Since the interactions are all
associated with the end of one open string joining the beginning of another
(as is clear by the Chan-Paton rules and the fact that the resulting action
is a single trace), every time two fields appear multiplied in the 
D-brane action, such
an operator will appear.  This is exactly the action of gauge theory
on the noncommutative torus.  

The essential phenomenon can already be seen in the D$0$-brane theory
without T-duality.
Consider a state containing an open string starting on D$0$-brane $i$,
winding $w_1$ times around $x^1$ and ending on brane $j$.
If one carries it about the loop
$x^2$, it will pick up a phase $e^{2\pi iB}$.  Thus its interactions with
the strings starting on $j$, winding $w_2$ times around $x^2$ and ending on
$i$ must contain a relative phase
$e^{2\pi iBw_1w_2}$ between the case of attaching at $i$ and attaching at $j$.

To come back to 
the original issue of minimal length, we have found that 
after dualizing to a description with $B=0$,
the radii of the
dual torus on which the final $2+1$-dimensional theory is defined are
$1/R_1$ and $1/R_2$, and do become large in the limit.  

\medskip

A context in which the nature of the limit
\tlimit\ is important is the recent
proposal by Sen and by Seiberg \senseiberg\ for 
a general definition of Matrix theory \refs{\bfss,\susslc}.
They propose quite generally that Matrix theory definitions of M theory
and DLCQ M theory (with a compact null direction)
can be obtained from type \IIa\ superstring backgrounds by considering
a sector with D$0$-branes and taking a scaling limit, which includes
the limit $R_1,R_2\rightarrow 0$.

Recently in \cds\ it was proposed that DLCQ M theory backgrounds with
non-zero background three-form $C_{-ij}$ along the null direction have
Matrix theory definitions obtained by replacing conventional gauge theory
with gauge theory on the noncommuting torus.
By the standards of superstring duality, quite strong evidence (equivalence
of the BPS mass formulas on both sides) was given.  On the other hand
gauge theory on the noncommuting torus has been unjustly neglected by
physicists and basic questions such as whether it is a well-defined quantum
theory have not yet been answered, leaving some room for scepticism.

Applying the idea of \senseiberg, we can define Matrix theory in these
backgrounds by starting with type \IIa\ superstring theory with non-zero
$B_{ij}$ field and taking a scaling limit.  This corresponds to M theory
$\int d^{11}x \ C_{11,ij}$ which under a boost turns directly into the
background of interest.  Thus the result we just described leads directly
to matrix theory on the noncommutative torus as proposed in \cds.

We see that gauge theory on the noncommutative torus is a well-defined
quantum theory, considered as a subsector of string theory.
This argument might well prove that its low energy limit
decoupled from string theory makes sense as a renormalizable
quantum theory.  The only caveat is that the non-locality of the theory
might invalidate the standard effective field theory arguments.
Since these are concrete theories with an action and a conventional
perturbative expansion, this question can be studied using standard techniques,
and we look forward to a detailed investigation of this point.

\listrefs
\end